\documentclass[aps,pre,twocolumn,showpacs,epsfig,epsf,amssymb]{revtex4}
\usepackage{graphicx}

\begin{document}

\title{Electrophoretic Properties of Highly Charged Colloids: A Hybrid MD/LB Simulation Study}

\author{Apratim Chatterji$^{1, 2}$, J\"urgen Horbach$^{1}$}

\affiliation{$^{1}$Institut f\"ur Physik,
Johannes--Gutenberg--Universit\"at, Staudinger Weg 7, D--55099 Mainz, Germany\\
$^{2}$Department of Physics, University of Toronto, 60 St. George Street,
Toronto M5S--1A7, Canada}

\begin{abstract}
Using computer simulations, the electrophoretic motion of a positively charged
colloid (macroion) in an electrolyte solution is studied in the framework
of the primitive model. In this model, the electrolyte is considered as
a system of negatively and positively charged microions (counterions
and coions, respectively) that are immersed into a structureless
medium. Hydrodynamic interactions are fully taken into account by applying
a hybrid simulation scheme, where the charged ions (i.e.~macroion and
electrolyte), propagated via molecular dynamics (MD), are coupled to a
Lattice Boltzmann (LB) fluid.  In a recent electrophoretic experiment by
Martin-Molina {\it et al.} [J. Phys. Chem. B {\bf 106}, 6881 (2002)],
it was shown that, for multivalent salt ions, the mobility $\mu$ initially
increases with charge density $\sigma$, reaches a maximum and then
decreases with further increase of $\sigma$.  The aim of the present
work is to elucidate the behaviour of $\mu$ at high values of $\sigma$.
Even for the case of monovalent microions, we find a decrease of $\mu$ with 
$\sigma$. A dynamic Stern layer is defined that includes all the counterions 
that move with the macroion while subject to an external electrical field. 
The number of counterions in the Stern layer, $q_0$, is a crucial parameter 
for the behavior of $\mu$ at high values of $\sigma$. In this case, the mobility 
$\mu$ depends primarily on the ratio $q_0/Q$ (with $Q$ the valency of the macroion).
The previous contention that the increase in the distortion of the electric double layer
(EDL) with increasing $\sigma$ leads to the lowering of $\mu$ does not hold
for high $\sigma$. In fact, we show that the deformation of the EDL 
decreases with increase of $\sigma$.
The role of hydrodynamic interactions is inferred from direct comparisons to
Langevin simulations where the coupling to the LB fluid is switched off.
Moreover, systems with divalent counterions are considered. In this case,
at high values of $\sigma$ the phenomenon of charge inversion is found.
\end{abstract}


\maketitle

\section{Introduction}

The electrophoretic motion of charged macroions in an electric
field ${\bf E}$ is a longstanding issue in colloid science
\cite{lyklema1,lyklema,hunter,olivares,saville,wowt,brien,oshina,booth,ovbeek,tata}.
A complex interplay occurs between hydrodynamic, electrostatic, and
thermal forces. A macroion of charge $Q$ drags along an electric
double layer (EDL) of small counterions with it and the number
of charges ``bound'' in the EDL depends on the charge density,
surface properties, ion concentration and liphophility of the
macroion as well as the specific properties of counterions and salt
ions~\cite{lemay}. Since it is difficult to experimentally probe and
control these different parameters at microscopic length and time
scales, the interpretation of experimental results often remains
unclear.  There is, e.g., no clear understanding of the variation of
the mobility $\mu$ with the surface charge density $\sigma$ of the
macroion \cite{lobaskin,lobaskin1,alvarez}.

Most of the theoretical studies on electrokinetic phenomena consider the
Coulomb interactions between macroions on the level of the linearized
Poisson--Boltzmann equation \cite{hunter,wowt,brien}.  This equation,
valid for weakly charged macroions and low salt concentration, describes
the formation of an EDL. It results in a screened Coulomb potential
(Yukawa potential) as the effective interaction potential between the
macroions. The central parameter of this potential is the Debye screening
length $\lambda_{\rm D}\equiv 1/\kappa$ (with $\kappa$ the screening
parameter) that measures the spatial extent of the EDL. When a charged
colloid moves in the presence of an electric field, the spherical EDL is
distorted and it is coupled to the hydrodynamic flow of the solvent. In
theoretical approaches \cite{olivares,torrie,netz,bocquet,alvarez},
hydrodynamic effects are incorporated by coupling the Poisson equation
for the electrostatics to the Navier--Stokes equations. In the
resulting electrokinetic equations \cite{saville}, so--called $\zeta$
potential is a central quantity. It is defined as the electrostatic
potential $\psi(z_s)$ at a distance $z_s$ from the colloidal surface
where the fluid around the colloid is at rest with respect to the
colloid motion. The imaginary surface at distance $z_s$ from the
colloidal surface is called surface of shear or slipping surface
\cite{saville,olivares,netz,bocquet}. 
 Some of the counterions that are between
the surface of the macroion and the surface of shear are often assumed
to move along with the macroion and define the so--called {\it dynamic}
Stern layer \cite{saville,netz}.

The $\zeta$ potential can be related to the electrophoretic mobility
$\mu=V_{\rm M}/E$ (with $V_{\rm M}$ the steady--state macroion
velocity). In the Helmholtz--Smoluchowski limit, i.e.~for $\kappa R_{\rm
M} \rightarrow \infty$ (with $R_{\rm M}$ the radius of the macroion),
the mobility of the macroion can be obtained from the $\zeta$ potential
via $\mu = \epsilon \zeta(z_s)/\eta$ \cite{lyklema,alvarez,bocquet},
where $\epsilon$ is the permittivity of the surrounding medium
and $\eta$ is the shear viscosity of the fluid.  In the opposite
H\"uckel--Onsager limit, $\lambda_D >> R_{\rm M}$, the mobility is
given by $\mu = 2 \epsilon \zeta(z_s)/ 3 \eta$ \cite{hunter}. The
definition of the potential $\zeta(z_s)$ at the surface of shear is
usually based on no--slip boundary conditions at the colloidal surface
\cite{bocquet}. Since $\kappa R_{\rm M} \rightarrow \infty$ implies
a very small Debye length, the distance of the surface of shear from
the colloidal surface would be typically in the nanometer range. At
such length scales the validity of no--slip boundary condition is
not clear (although a hydrodynamic description might be still valid
\cite{jh06}). Calculations of $\mu$ which cover the entire range
$\kappa R_{\rm M}$ from the Helmholtz--Smoluchowski limit to the
H\"uckel--Onsager limit have been done by Henry \cite{hunter}. But
these calculations, valid for low $\zeta$, assumed that the counterion
charge density is unaffected by the applied field.

O'Brien and White \cite{brien} did extensive numerical calculations
to calculate $\mu$ versus $\zeta$ for different values of $\kappa
R_{\rm M}$ in the framework of electrokinetic equations. For $\kappa
R_M > 3$, they obtained a pronounced maximum when plotting $\mu$ as a
function of $\zeta$. This behaviour is attributed to the competition
between the driving force due to $E$ and the retarding forces due to
the distortion of the charge cloud constituting the EDL. The driving
force increases linearly with $\zeta$ (which is proportional to $Q$),
whereas the retarding force is proportional to $\zeta^2$. Thus for
high values of $\zeta$, $\mu$ decreases with $\zeta$ because then the
retarding force dominates. Moreover, O'Brien and White found that for
multivalent ions, the position of the maximum shifts to lower values
of $\zeta$. They argue that the distortion of the double layer is
increased by multivalent counterions resulting in the increase of
retardation forces against the motion of the colloid. Qualitatively
similar results were also obtained by the non--linear solution by Oshina
{\it et al.}~\cite{oshina} and by perturbation expansion in powers of
$\zeta$ by Booth~\cite{booth} and Overbeek~\cite{ovbeek}. There are also
other models of more microscopic origin which look at charged systems,
but these models focus mainly on the structure and dynamics of the
dynamic Stern layer in planar geometry~\cite{bocquet,netz} rather than
on electrophoretic properties of charged spherical colloids.  Moreover,
some of these studies disregard hydrodynamic interactions \cite{netz}
or consider only zero--temperature properties \cite{patra}.

Electrophoresis experiments on colloidal suspensions
\cite{alvarez,lemay,tp1,tp2,tp3,tp4,shapran05,grosberg} allow to
determine accurately the electrophoretic mobility $\mu$ as a function of
different parameters such as $\kappa R_{\rm M}$, surface charge density
of colloids, etc.  However, it is difficult to disentangle hydrodynamic
effects from those due to electrostatic interaction. Moreover, in order
to fit experimental data to theoretical models, ``renormalized charges''
of the colloids have to be introduced \cite{shapran05}. 

In a recent experimental study by Martin--Molina {\it et
al.}~\cite{alvarez}, a maximum was found in the electrophoretic
mobility $\mu$, plotted as a function of the macroion's surface charge
density $\sigma$ (note that there is a linear relationship between
$\sigma$ and the $\zeta$ potential in the framework of the linearized
Poisson--Boltzmann equation). The latter maximum occurs in the case of a
2:1 (counterion-charge:coion-charge) electrolyte. A similar result can be
also found in the numerical paper by O'Brien and White \cite{brien} for
a 2:1 electrolyte. However, the experiments were performed for systems
of highly charged macroion's. 
In this case, it is not clear whether the calculations by O'Brien and
White yield a correct description, even on a qualitative level. In this
work, we aim at addressing this issue by using computer simulation
techniques that may elucidate electrophoresis with highly charged
particles on a microscopic level.

Molecular dynamics (MD) simulations of colloidal systems suffer from
the large separation in length and time scales of solvent and colloidal
particles.  The problem of simulating millions of solvent particles to
account for hydrodynamic effects can, however, be circumvented by using
simulation techniques such as the Lattice Boltzmann (LB) \cite{lbe}
method or other Navier--Stokes equation solver (see, e.g.~\cite{kim06})
to model the solvent.  In this work, we employ a hybrid LB/MD
method \cite{ac,ac2} to investigate the electrophoretic motion of a highly
charged macroion in an electrolyte solution, thereby modeling macroion
and electrolyte in the framework of the so--called primitive model (in a
similar way as in Ref.~\cite{kreer06}). Our aim in particular
is to investigate those structural and dynamical aspects of the EDL 
which are out of scope of mesoscopic theories.

In the following, we are interested in the case of high values of
$\sigma$, i.e.~the regime beyond the aforementioned maximum in the
$\mu-\sigma$ curve. In particular the counterion distribution around
the macroion is measured to provide insight into the electrical retarding
forces affecting the mobility. In order to disentangle hydrodynamic
from electrostatic effects, the hydrodynamic medium (LB fluid) is also
switched off, thus simulating the system via Langevin dynamics. Systems
with monovalent and divalent salt ions are considered. In the latter
case, the phenomenon of charge inversion is observed.

The rest of the paper is organized as follows: We describe the LB/MD
method in Sec.~II.  In Sec.~IIIA, we present the results for the
monovalent electrolyte solution (microions) and in Sec.~IIIB, we focus
on the charge inversion phenomenon with divalent microions.  Finally,
conclusions are given in Sec.~IV.

\section{Model and details of the simulation}
The electrophoretic motion of a macroion in an electrolyte solution is
studied in the framework of the so--called primitive model. We consider
a system of a macroion of charge $Q=Z_{\rm M}e$ (mass $M=60$ a.u.) and
microions of charge $Z_{\rm ct}=-1 e$ (counterions) and $Z_{\rm co}=1 e$
(coions), each of which are of mass 4 a.u.. The interaction potential
between a particle of type $\alpha$ and a particle of type $\beta$
($\alpha, \beta = {\rm M, ct, co}$) separated by a distance $r$ from
each other is given by
\begin{equation}
  u_{\alpha \beta} =
  \frac{Z_{\alpha} Z_{\beta} e^2}{4 \pi \epsilon r}
  + A_{\alpha \beta}
    \exp \left\{ - B_{\alpha \beta} (r - \sigma_{\alpha \beta} ) \right\}
  \label{eq_pot}
\end{equation}
where $e$ is the elementary charge and $\epsilon$ the dielectric constant.
We choose the value $\epsilon=80 \epsilon_0$ (with $\epsilon_0$ the
vacuum dielectric constant) for water at room temperature.  The parameters
$\sigma_{\alpha \beta}$ denote the distance between two ions at contact,
$\sigma_{\alpha \beta}= R_{\alpha} + R_{\beta}$, where $R_{\alpha}$
is the radius of an ion of type $\alpha$. 
In the following, we use values for $R_{\rm M}$ that vary from 10\,\AA~to 20\,\AA. 
For the microions, we set $R_{\rm ct}=R_{\rm co}=1$~\AA.  
The exponential in Eq.~(\ref{eq_pot}) is an approximation to a hard sphere interaction
for two ions at contact. For the parameters $A_{\alpha \beta}$ we choose
$A_{\rm MM}=1.84$~eV, $A_{\rm M,ct}=A_{\rm M,co}=0.05565$~eV, and
$A_{\rm ct,ct}=A_{\rm ct,co}=A_{\rm co,co}=0.0051$~eV. The parameters
$B_{\alpha \beta}$ are all set to $4.0$~\AA$^{-1}$.  The long--ranged
Coulomb part of the potential and the forces were computed by Ewald sums
in which we chose $\alpha_E=0.05$ for the constant and a cutoff wavenumber
$k_{\rm c}=2\pi\sqrt{66}/L$ in the Fourier part~\cite{allen,frenkel}.
The linear dimension $L$ of the simulation box is $L=160$\,\AA, using
periodic boundary conditions. All simulations were done at the temperature
$T=297$\,K.

Now, the crucial step is to model the solvent. To a first approximation,
the effect of the solvent on colloidal particles (macroions) can be
described by a Langevin equation where one assumes that on the typical
time scale of the colloidal particles their collisions with solvent
particles are due to Gaussian random forces ${\bf f}_{{\rm
r},i}$. These forces lead to a systematic friction force $-\xi_0 {\bf
V}_i(t)$ on the colloids, where $\xi_0$ is the friction coefficient and
${\bf V}_i(t)$ the velocity of particle $i$ at time $t$. The resulting
equations of motion are
\begin{equation}
  \label{langevin}
  M \frac{d^2 {\bf R}_i}{dt^2} = {\bf F}_{{\rm c}, i}
                 -\xi_0 {\bf V}_i(t)
                 + {\bf f}_{{\rm r},i} \, ,
\end{equation}
where ${\bf R}_i$ and ${\bf V}_i(t)$ denote respectively position and velocity
of a colloidal particle $i$ ($i=1, ... , N$ with $N$ the total number of colloids) and
${\bf F}_{{\rm c}, i}$ is the conservative force acting on the particle. 
The Cartesian components of the random forces, $f_{{\rm
r},i}^{\alpha}$ ($\alpha= x, y, z$), are uncorrelated random numbers
with zero mean, i.e.
\begin{eqnarray}
 \left< f_{{\rm r},i}^{\alpha} ({\bf R}_i,t) \right> & = & 0 \label{fr1} \\
 \left< f_{{\rm r},i}^{\alpha}({\bf R}_i,t)
        f_{{\rm r},i}^{\beta}({\bf R}_i^{\prime},t^{\prime}) \right> & = &
        A \delta_{\alpha \beta} \delta( {\bf R}_i - {\bf R}_i^{\prime})
        \delta(t - t^{\prime}) \, . \label{fr2}
\end{eqnarray}
The amplitude $A$ is determined by the fluctuation--dissipation theorem,
$A=2k_BT\xi_0$. In our model, the microions are also Brownian particles
and in thermal equilibrium with the solvent.

Equations (\ref{langevin})--(\ref{fr2}) still do not provide a complete
description of the dynamics in colloidal suspensions. In these equations
the mass and momentum transport by the solvent is ignored that leads to
the hydrodynamic interactions between the colloidal particles. However,
hydrodynamic interactions can be incorporated rather easily into the
Langevin description by replacing the absolute velocity ${\bf V}_i(t)$ of
particle $i$ in Eq.~(\ref{langevin}) by its velocity relative to the fluid
velocity field ${\bf u}({\bf R}_i,t)$ at the position of the particle
${\bf R}_i$.  The ``hydrodynamic force'' on particle $i$ is then given by
\begin{equation}
  \label{fric}
  {\bf F}_{{\rm hyd},i} = - \xi_0 \left[{\bf V}_i(t) - {\bf u}({\bf R}_i,t) \right]
                          + {\bf f}_{{\rm r},i}
\end{equation}
In a simulation, the velocity field ${\bf u}({\bf R}_i,t)$ can be calculated
from any Navier--Stokes equation solver, whereby thermal fluctuations
have to be included in the framework of fluctuating hydrodynamics.
In this work, we use a Lattice--Boltzmann (LB) \cite{lbe,ladd} scheme
to compute ${\bf u}({\bf R}_i,t)$. Since the LB method yields the
velocity field ${\bf u}$ on a lattice, an interpolation scheme has to
be used to determine ${\bf u}$ at the position of the particle ${\bf
R}_i$. Very recently, Ahlrichs and D\"unweg~\cite{ahlrichs} have proposed
a hybrid MD/LB scheme on the basis of the frictional coupling force,
Eq.~(\ref{fric}), to simulate polymers in solution.

Up to now, we have considered the colloidal particles as point particles.
However, real colloids are extended objects with rotational degrees
of freedom. The problem is to ``implant'' an extended object such as
a sphere of radius $R_{\rm H}$ into a LB fluid. To this end, we follow Ladd's
method \cite{ladd} and represent the sphere (or any other object) by uniformly
distributed boundary points on its surface, where the surface is
permeable for the LB fluid. We have recently proposed a sphere model
with 66 boundary nodes of mass $M/66$ with $M$ being the total mass
of the colloidal particle \cite{ac}. This model for a spherical particle is also
considered in the following. Each of the boundary points is coupled to
the LB fluid by the force given by Eq.~(\ref{fric}). The total force on
the particle is determined by the sum over the forces on the boundary
nodes (of course, according to Newton's third law, these forces with 
opposite sign are given back to the 
LB fluid). From this total force the torque on the particle is calculated,
and the total force and torque are then used to update the translational
and rotational velocity of the particle, respectively. More details on
our MD/LB scheme can be found elsewhere~\cite{ac}.

For the LB fluid a standard D3Q18 model was used, with which we solved
the linearized Navier--Stokes equations (the details of the D3Q18 model
can be found in review articles and the book by Succi~\cite{lbe,ladd}).
We consider an incompressible fluid in the creeping flow limit (zero
Reynolds number) in our studies.  The kinematic viscosity was set to
$\nu=0.0238 a^2/\tau$ with $a$ the lattice constant and $\tau$ the
elementary time unit of the LB fluid. The density of the LB fluid was
set to $\rho=1.0 m_0/a^3$ ($m_0$: mass unit of the LB fluid). Thermal
fluctuations were introduced via the addition of Gaussian random numbers
to the stress tensor as proposed by Ladd~\cite{ladd}. The LB fluid is
modelled on a cubic lattice with $40^3$ lattice nodes, thus the lattice
constant is $a=4$\,\AA~(for the linear dimension $L=160$\,\AA~of the
simulation box).  The counterions and coions are considered as point
particles with respect to the interaction with the LB fluid. The 66
boundary nodes of the macroion are placed on a sphere of radius 
$R_{\rm H}=10$\,\AA. 
The effective hydrodynamic radius and the viscous retarding force on
the macroion is thus determined by $R_{\rm H}$. On the other hand, the
variation of $R_{\rm M}$ from 10\,\AA~to 20\,\AA~(see next section)
allows a change of the macroion's surface charge $\sigma$ without
changing the hydrodynamic coupling of the particle to the LB fluid.
For the friction constant $\xi_0$ in
the coupling force, Eq.~(\ref{fric}), a total value of $6.6 m_0/\tau$
is assigned to the macroion which corresponds to $\xi_0=0.1 m_0/\tau$
for each of the 66 boundary nodes.  In a recent publication~\cite{ac},
we have shown that this value of $\xi_0$ corresponds to nearly stick
boundary conditions. 
For the microions, a different value for the friction coefficient
is used, denoted by $\xi_{0b}$ in the following. Below we discuss
the influence of $\xi_{0b}$ on the electrophoretic mobility (see
Fig.~\ref{fig6}). Unless otherwise noted, the value $\xi_{0b}=0.025
m_0/\tau$ was chosen. With respect to the LB fluid, the microions are
treated as point particles, but we remind the reader that their Coulomb
radii are set to $R_{\rm ct}=R_{\rm co}=1$\,\AA.

%
\begin{table}
\begin{center}
\begin{tabular}{|c|c|c|c|c|}
\hline $Q$ [$e$] & $N_{\rm ct}$ & $N_{\rm co}$ & $\kappa$ [\AA$^{-1}$] & $ \kappa R_M$  \\ \hline \hline
121 & 471 & 350 & 0.127 & 2.54\\ \hline
255 & 555 & 300 & 0.133 & 2.66 \\ \hline
351 & 651 & 300 & 0.137 & 2.74 \\ \hline
401 & 651 & 250 & 0.126 & 2.52 \\ \hline
601 & 801 & 200 & 0.141 & 2.82 \\ \hline
801 & 1001 & 200 & 0.154 & 3.08 \\ \hline
\end{tabular}
\end{center}
\caption{\label{tab1} Charge of the macroion $Q$,
number of counterions $N_{\rm ct}$, number of coions $N_{\rm co}$,
and the corresponding values of the screening parameter $\kappa$ and the dimensionless
quantity $\kappa R_M$ that were
used in the simulations for $R_{\rm M}=20$\,\AA.}
\end{table}
We have done MD/LB simulations of a single macroion of charge $Q$
in an electric field $E_x$ pointing in the positive $x$ direction. The
charge $Q$ was varied from $121\,e$ to $801\,e$. The number of
counterions and coions ($N_{\rm ct}$ and $N_{\rm co}$, respectively) used
for a given value of $Q$ are listed in Table~\ref{tab1}. Also
included in this list is the Debye screening parameter $\kappa=[4
\pi \lambda_B (n_{\rm ct}+n_{\rm co})/L^3]^{1/2}$ which was roughly
kept constant around 0.13\,\AA$^{-1}$ (the Bjerrum length $\lambda_{\rm
B}=e^2/(4 \pi \epsilon_{\rm r} \epsilon_0 k_B T)$ is in our case equal to
about $7$~\AA).  For comparison, we also carried out simulations with a
``Langevin dynamics'' (LD) where the coupling to the LB fluid was switched
off, i.e.~${\bf u}({\bf R},t)=0$.  For the LB/MD and LD simulations, the
equations of motion were integrated by a Heun algorithm with a time step
of $1$\,fs.  This very small time step is necessary because we consider
explicitly counterions and coions as microscopic particles.

The simulations were done as follows: We first equilibrated the system
for 25000 time steps without electric field and without coupling to the
LB fluid.  Then, the system was coupled to the LB fluid and the electric
field was switched on, followed by simulation runs over 400000 time steps.
After 100000 time steps, the steady state was reached and the positions
and velocities of the ions were stored every 500 time steps to determine
the averaged quantities such as the steady state velocity $V_{\rm M}$
of the macroion.

\section{Results}
Now the simulation results are presented for the steady--state
electrophoretic motion of a macroion in an electrolyte. Systems of
highly charged macroions are considered, i.e.~their surface charge
densities vary between $\sigma=0.02\,e/$\AA$^2$ (38 $\mu$C/cm$^2$)
and $\sigma=0.3\,e/$\AA$^2$ (500 $\mu$C/cm$^2$). The density of
coions ($N_{\rm co}=300$) in our simulation box corresponds to a salt
concentration of $0.012$\,mol/l.

\subsection{Monovalent microions}
In this section, we consider electrolyte solutions that consist of
monovalent counter-- and coions.  For a microscopic understanding of
electrophoresis, it is of particular interest to display the {\it dynamic}
distribution of counterions in the vicinity of the macroion, i.e.~in
the EDL. This distribution is determined by an interplay between the
electrostatic attraction and the hydrodynamic flow around the macroion.
Some of the counterions in the EDL move along the same direction as the
macroion, whereas, due to the electric field, other counterions in the
EDL are accelerated in the opposite direction. We analyze the dynamic
behavior of the counterions by their average velocity $V_{\rm ct}(r)$
as a function of the radial distance $r$ from the center of the macroion
and the cumulative counterion charge $q(r)$ around the macroion.

\begin{figure}[tb]
\centering
\includegraphics*[width=0.8\columnwidth]{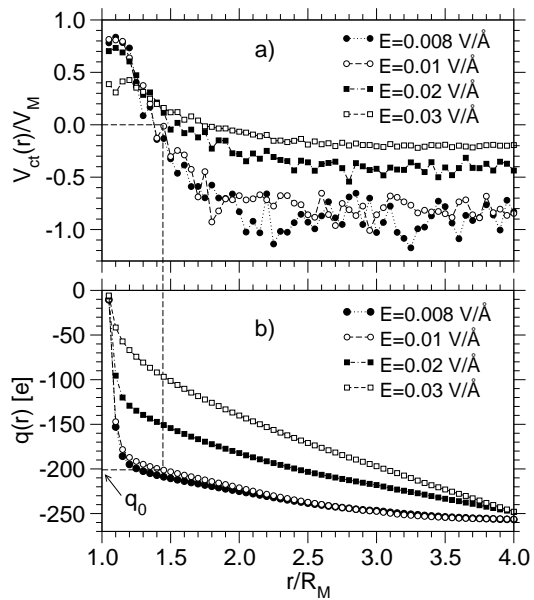}
\caption{\label{fig1}
a) $V_{\rm ct}(r)/V_{\rm M}$ as a function of $r/R_{\rm M}$ for different
values of the electric field, as indicated.  Data is shown for 1 macroion
of charge $Q=255 e$, $N_{\rm ct} = 555$ monovalent counterions and $N_{\rm
co}=300$ monovalent coions. b) Cumulative counterion charge $q(r)$ for the
same parameters as in a).  $q_0$ is the cumulative counterion charge at
which $V_{\rm ct}(r)$ is zero.  The determination of $q_0$ is indicated
in the figure by the dashed lines for the example $E=0.01$ V/\AA.
}
\end{figure}
First, we consider a system consisting of one macroion of charge $Q=255
e$, $N_{\rm ct} = 555$ monovalent counterions and $N_{\rm co}=300$
monovalent coions. For this case, Fig.~\ref{fig1} displays $V_{\rm
ct}(r)/V_{\rm M}$ (with $V_{\rm M}$ the steady state macroion velocity)
and $q(r)$ for different choices of the electric field $E$.  Within the
statistical errors, the results essentially coincide for the two lowest
values of the electric field, $E=0.01$\,V/\AA~and $E=0.008$\,V/\AA.
The choice $E=0.01$\,V/\AA~is used for all further results presented
in this work, in particular for the estimates of the electrophoretic
mobility, $\mu = V_{\rm M}/E$. 
As shown previously \cite{ac}, for $Q\ge255e$ the linear response regime
is essentially achieved for $E\le 0.01$\,V/\AA.

From the behavior of $V_{\rm ct}(r)/V_{\rm M}$, different regions can
be identified with respect to the distance from the center of the
macroion. Close to the macroion's surface a layer with a thickness
of about $0.25\,R_{\rm M}$ is formed where $V_{\rm ct}/V_{\rm M}$ is
constant with a value around $0.8$ for the two lowest values of $E$.  In the
following, this region is called the dynamic Stern layer, where, due to the
electrostatic attraction by the macroion, counter\-ions are essentially
condensed onto the surface of the macroion.  Beyond the Stern layer, the
counterion velocity changes quickly its sign, thus indicating a motion
in the direction opposite to that of the macroion. The point where the
counterion velocity vanishes can be used as a measure of the extent of
the Stern layer. As shown in Fig.~\ref{fig1}, the cumulative counterion
charge at this point, $q_0$, is significantly smaller than the bare charge
of the macroion (e.g.~$q_0\approx -200\,e$ for $E=0.01$\,V/\AA). Not until
$r$ is of the order of 3--4\,$R_{\rm M}$, the counterion charge $q(r)$ is
equal to the macroion's charge, thus completely neutralizing the latter.
For high values of $E$, e.g.~$E=0.03$\,V/\AA, the counterions are stripped
off the surface of the macroion, since the force due to the electric field
dominates over the Coulomb attraction between macroion and counterions.
This leads to a lower value of $q_0$ for high values of $E$ and a less
efficient shielding of the macroion compared to the case of low values
of $E$. This in turn increases the velocity $V_{\rm M}$ of the macroion
and thus explains the low values of $V_{\rm ct}/V_{\rm M}$ for large
values of $E$ at distances far from the colloidal surface.

\begin{figure}[tb]
\centering
\includegraphics[width=0.8\columnwidth]{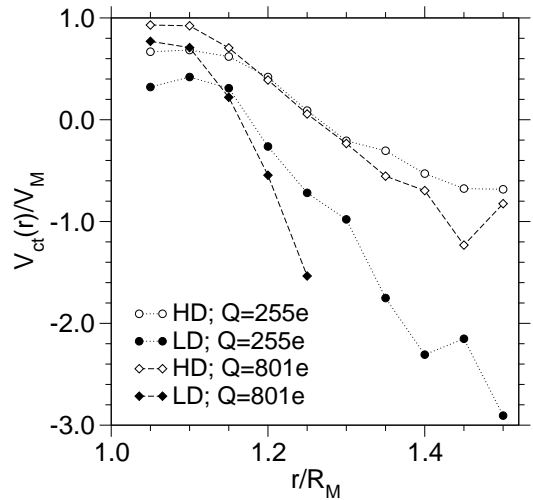}
\caption{\label{fig2}
$V_{\rm ct}(r)/V_{\rm M}$ as a function of $r/R_{\rm M}$ for two different
values of the bare charge $Q$, as indicated. For the HD case, the data
from Fig.~\ref{fig1} for $Q=255$\,e and $E=0.01$\,V/\AA~are plotted
(open circles).  The closed circles are the corresponding results from LD
simulations. Also shown are results for $Q=801$ (open and closed diamonds
for HD and LD, respectively). In this case, the system contains $N_{\rm
ct}=1001$ counterions and $N_{\rm co}=200$ coions.
}
\end{figure}
One might expect, that the region between the Stern layer and the
point where $q \approx Q$, is strongly affected by hydrodynamic flow
features. That this is indeed the case can be infered from a comparison
to LD simulations where the coupling to the LB fluid is switched off
(in the following, we refer to simulations with a coupling to LB 
as HD simulations).

\begin{figure}[t]
\centering
\includegraphics[width=0.8\columnwidth]{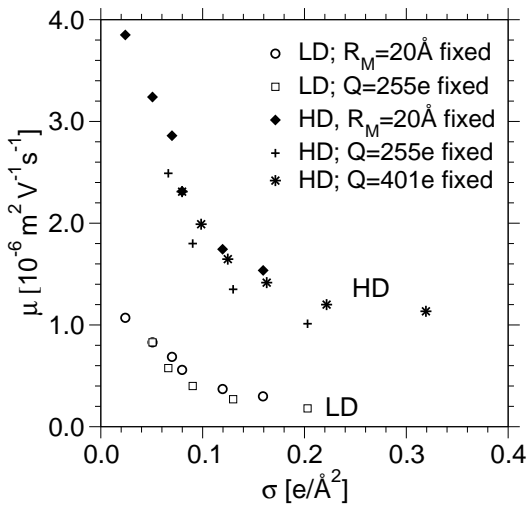}
\caption{\label{fig3}
The macroion mobility $\mu$ as a function of the surface charge density
$\sigma$ for LD and HD simulations as indicated.  The charge density
$\sigma$ is either varied by changing the radius $R_{\rm M}$ of the
macroion from $10$\,\AA to $20$\,\AA~keeping $Q$ fixed at $Q=255\,e$
or $Q=401\,e$, or by changing $Q$ from $Q=121\,e$ to $Q=801\,e$
keeping $R_{\rm M}$ fixed at $20$\,\AA.  The number of counterions
and coions used for each value of $Q$ is mentioned in text.  For all
data, the ``hydrodynamic radius'' is chosen to be constant at $R_{\rm
H}=2.5\,a=10$\,\AA.}
\end{figure}
In Fig.~\ref{fig2}, $V_{\rm ct}(r)/V_{\rm M}$ from LD simulations is
compared to the same quantity from HD simulations for the two macroion
charges $Q=255\,e$ and $Q=801\,e$ (in the latter case the system contains
$N_{\rm ct}=1001$ counterions and $N_{\rm co}=200$ coions).  In the HD
case, we see that for $Q=801\,e$, the ratio $V_{\rm ct}(r)/V_{\rm M}$
is very close to one in the Stern layer region. However, for $Q=255\,e$
the ratio is significantly smaller. The strong Coulomb attraction
dominates the  viscous drag and thermal fluctuations for high charges,
such that layers of counterions nearest to the colloidal surface nearly
stick to the surface. But for lower macroion charges, one starts to see
deviations from the assumption that the {\it dynamic} Stern layer consists
of immobile counterions \cite{lyklema,bocquet}. Moreover, the Stern
layer is extended, and it is not restricted to one layer of microions
closest to the macroion surface as observed in \cite{bocquet}. This is
due to the much higher values of the surface charge density $\sigma$
used in our study.

\begin{figure}[t]
\vspace*{0.2cm}
\centering
\includegraphics[width=0.8\columnwidth]{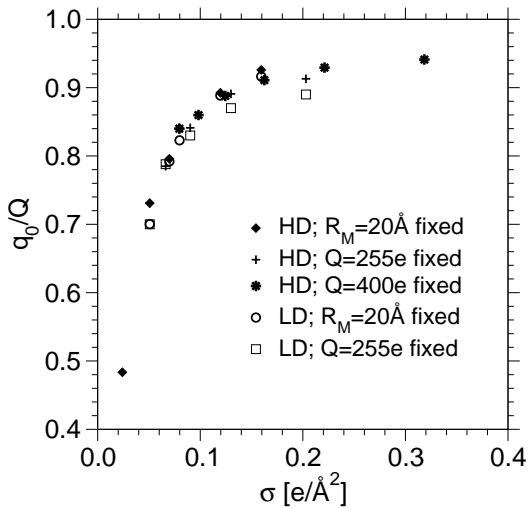}
\caption{\label{fig4}
Ratio $q_0/Q$ as a function of $\sigma$.  As indicated, in the different
data sets either $Q$ or $R_{\rm M}$ are fixed while $\sigma$ is varied
(see also Fig.~\ref{fig3}).}
\end{figure}

As indicated by Fig.~\ref{fig2}, in the LD case the motion of the
counterions is less correlated to that of the macroion than in the
HD case.  The value of $V_{\rm ct}(r)/V_{\rm M}$ is smaller in the
Stern layer region than in the corresponding HD data.  Moreover,
the counterions reverse their velocity at about $0.15\, R_{\rm M}$
away from the macroion surface, followed by a more rapid decrease of
$V_{\rm ct}(r)/V_{\rm M}$ than in HD. From data presented below (in
Fig.~\ref{fig4}), we will see that the number of counterions carried
along in the Stern layer is almost the same for LD and HD, which reveals
that, in the LD case, the counterions are more densely packed in the
Stern layer.

In order to investigate the mobility $\mu$ of the macroion as a function
of its surface charge density $\sigma = Q/(4 \pi R_{\rm M}^2)$, two
different types of simulations were performed. First, the charge $Q$
was varied from $Q=121\,e$ to $801\,e$ while keeping the radii $R_{\rm
H}$ and $R_{\rm M}$ fixed at $10$\,\AA~and $20$\,\AA, respectively.
The number of counter-- and coions used for a given value of $Q$
are listed in Table \ref{tab1}. Also included in this list is the
Debye screening parameter $\kappa=[4 \pi \lambda_{\rm B} (N_{\rm
ct}+N_{\rm co})/L^3]^{1/2}$ which was roughly kept constant around
$0.13$\,\AA$^{-1}$.  Secondly, runs for $Q=255\,e$ and $Q=401\,e$
were done in which the charge density $\sigma$ was varied by choosing
different macroion radii 10\,\AA$\le R_{\rm M}\le$ 20\,\AA~in steps
of $2$\,\AA~or $2.5$\,\AA~(note that the radius $R_{\rm H}$ remains
fixed at 10\,\AA~in these runs). The results for the different runs
are shown in Fig.~\ref{fig3} where $\mu$ is plotted as a function of
$\sigma$ for both HD and LD runs. In both cases, the mobility $\mu$
decreases with increasing $\sigma$. This behavior is in qualitative
agreement with theoretical calculations for high values of the surface
charge, i.e.~$\sigma> 0.01\,e/{\rm \AA}^2$~(see Ref.~\cite{alvarez} and
references therein). Note that, at small $\sigma$ the opposite behavior
is observed, i.e.~an increase of $\mu$ with $\sigma$~\cite{alvarez}.

For monovalent microions, Fig.~\ref{fig3} demonstrates that at constant 
screening parameter $\kappa$ the mobility $\mu$ is controlled by the surface 
charge density of the macroion. Discrepancies that can
be seen in the plot might stem from the slight variation of $\kappa$
in the different runs (see Table \ref{tab1}). The LD runs that are
shown in the figure exhibit a similar qualitative behavior.  However,
compared to HD, the LD data is shifted towards lower values of $\mu$.
This is because the hydrodynamic flow field is switched off in the LD
simulations.  The coupling of the LB fluid to the motion of the macroion
leads to a backflow effect in the LB fluid which enhances the mobility
of the macroion.  This backflow effect yields also stronger correlations
between the motion of the counterions and that of the macroion.  Thus,
as shown in Fig.~\ref{fig2}, the reduced counterion velocity $V_{\rm
ct}(r)/V_{\rm M}$ decays slower in the HD case than in the LD case.

The correlations between counterions and macroions are further considered
in Fig.~\ref{fig4}. Here, $q_0/Q$ is plotted vs.~$\sigma$. As before,
we vary $\sigma$ by changing $R_{\rm M}$ keeping $Q$ fixed or by changing
$Q$ with $R_{\rm M}$ fixed at $20$\,\AA.  As we see, the different data
sets fall roughly onto one master curve (note that this is even true
for the LD data).  The functional behavior of $q_0/Q$ reflects the one
for $\mu$. At small values of $\sigma$, the ratio $q_0/Q$ is ``rapidly''
increasing (associated with a rapid decrease of $\mu$) and it seems to
saturate at high $\sigma$ (as $\mu$ does). Thus, the electrophoretic
mobility shows a saturation when $q_0/Q$ is approaching unity. In the
latter case, the electric field sees a ``particle'' with an effective
charge $Q_{\rm eff}=Q-q_0$.

\begin{figure}[b]
\centering
\includegraphics[width=0.9\columnwidth]{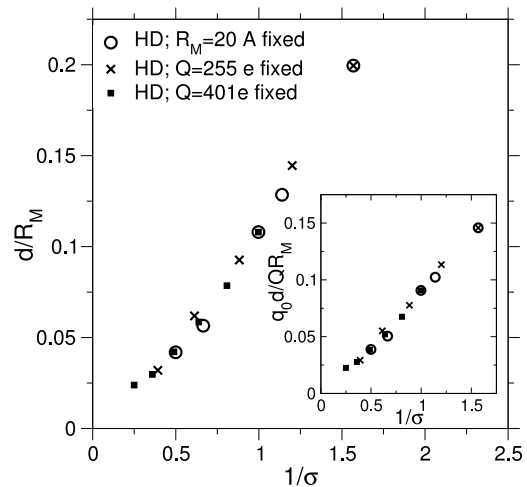}
\caption{\label{fig4a}
Plot of the measure of distortion $d$ of the Stern layer,
normalised by the Coulomb radius of colloid $R_{\rm M}$
as a function of the inverse surface charge density $1/\sigma$
(for the definition of `$d$' see text).
The inset shows $q_0 d/Q R_{\rm M}$ which 
is a measure of the charge separation due to the distortion
of the EDL.
}
\end{figure}
The regime of low macroion charges $Q$ has been studied recently
by Lobaskin {\it et al.}~\cite{lobaskin} using a similar LB/MD
technique. These authors considered a primitive model in the salt free
regime (i.e.~without coions). Their results agree with experimental 
data~\cite{lobaskin1} for low $Q$, where an increase of $\mu$ with $Q$
($\sigma$) is observed. In this regime, $\mu$ is controlled by the bare
charge $Q$ as there is no significant dynamic Stern layer to create a shielding
for $Q$. Hence the peak in $\mu$ observed in the experimental data
of Martin-Molina {\it et al.}~\cite{alvarez} denotes the point of crossover
from this regime to a regime where $\mu$ is controlled by $q_0/Q$. For
high values of $\sigma$, $\mu$ decreases with $\sigma$ due to greater
shielding for higher $Q$, whereas, for low values of $\sigma$ ($Q$)
the mobility increases with $\sigma$.

O'Brien and White \cite{brien} have proposed that the decrease in macroion
mobility can be attributed to the increasing distortion of the EDL at
high values of $\sigma$.  To see whether the distortion of the charge
cloud is indeed a relevant effect also at very high values of $\sigma$
(considered in this work), we quantify the distortion of the EDL as
follows: We compute the distance $d$ of the center of mass (CM) of the
EDL from the center of the macroion. Here, those counterions form
the EDL which are within the {\it dynamic} Stern layer and move along
the macroion. Figure \ref{fig4a} shows that the normalised distortion
$d/R_{\rm M}$ and the normalised quantity $q_0 d/Q R_{\rm M}$ decrease
with increasing $\sigma$. The quantity $q_0 d/Q R_{\rm M}$ is a measure
of the charge separation between the center of the macroion and center
of the counterion cloud. Thus, the amount of distortion becomes less
pronounced with increasing $\sigma$. This rules out the mechanism
proposed by O'Brien and White in the case of high values of $\sigma$.

\begin{figure}[t]
\centering
\includegraphics[width=0.8\columnwidth]{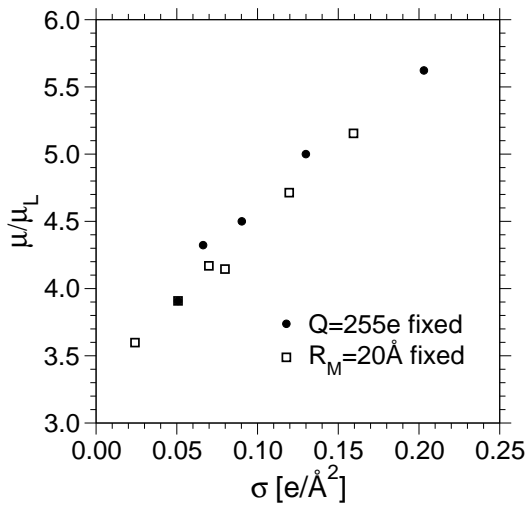}
\caption{\label{fig5}
Plot of $\mu_{\rm HD}/\mu_{\rm LD}$ versus the charge density
$\sigma$. The mobilities $\mu_{\rm HD}$ and $\mu_{\rm LD}$ correspond
to HD and LD runs, respectively.
}
\end{figure}
We have already infered from Fig.~\ref{fig2} that hydrodynamic backflow
due to the LB fluid enhances the correlations between the counterion and
the macroion motion as well as the absolute value of $\mu$. 
In Fig.~\ref{fig5}, the mobilities obtained from LD and HD runs (denoted by
$\mu_{\rm HD}$ and $\mu_{\rm LD}$, respectively) are directly compared
to each other by plotting the ratio $\mu_{\rm HD}/\mu_{\rm LD}$
as a function of $\sigma$. We see that $\mu_{\rm HD}/\mu_{\rm LD}$
increases almost linearly with $\sigma$. 

\begin{figure}[tb]
\centering
\includegraphics[width=0.8\columnwidth]{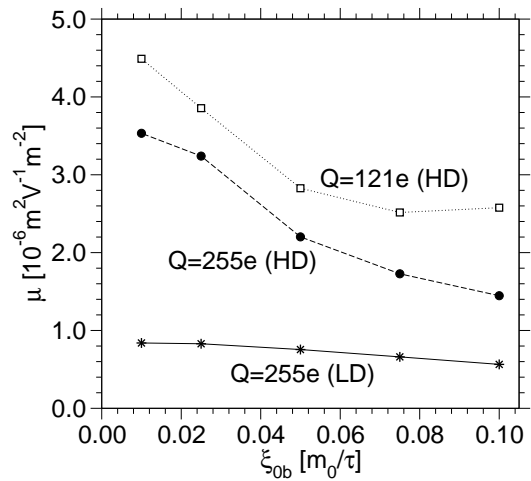}
\caption{\label{fig6}
Macroion mobility $\mu$ versus the counterion coupling constant $\xi_{0b}$
for LD and HD simulations. The colloidal friction $\xi_0$ is fixed at
$0.1 m_0/\tau$ and $R_{\rm M} = 20$ \AA. The lines are guides to the eye.
}
\end{figure}
Microscopically, we understand this by noting that the higher macroion
charge density leads to a higher number density of counterions per unit
volume in the moving EDL of the colloid. The counterion coupling to the
LB fluid results in an increased velocity of the LB fluid around the
macroion at higher densities. Thus the macroion feels less frictional
drag and, compared to the LD case, a higher value of $\mu$ is obtained.

In Fig.~\ref{fig6}, it is further explored how the counterion coupling
with the LB fluid affects the colloidal mobility. We keep the
friction coefficient for the macroion coupling to the LB fluid,
$\xi_0$, fixed and vary that for the microion coupling, $\xi_{0b}$
(for the definition see Sec. II). Whereas, for the HD simulations,
$\mu$ decreases significantly with $\xi_{0b}$, in the LD case, $\mu$
exhibits only a weak dependence on $\xi_{0b}$.  Thereby, the HD result
tends to approach the one from LD for high values of $\xi_{0b}$. In
the HD case, small values of $\xi_{0b}$ mean a weaker coupling of the
microion motion to that of the LB fluid, leading to an increase of the
macroion mobility.  The detailed understanding of this finding requires
further investigation.

Note that, in the LD simulation for $Q=121\,e$, a dynamic Stern layer
cannot be formed, because the Coulomb attraction is too weak in this
case. Hence, a high value of mobility results which is outside the range
used in Fig.~\ref{fig6}. On the other hand, in the HD simulation for
$Q=121\,e$ a well-defined dynamic Stern layer is seen and, as shown by
Fig.~\ref{fig6}, the behavior of $\mu$ is qualitatively similar to that
at $Q=255\,e$. This demonstrates the importance of the hydrodynamic
flow field for the behavior of the electrophoretic mobility of weakly
charged colloids.

\subsection{Divalent salt ions}
\begin{figure}[tb]
\centering
\includegraphics[width=0.8\columnwidth]{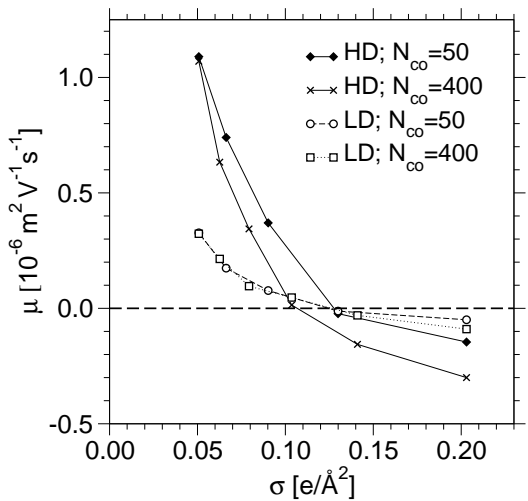}
\caption{\label{fig7}
Electrophoretic mobility $\mu$ as a function of surface charge density
$\sigma$ for two different systems using LD and HD, as indicated. The lines
serve as guides to the eye.}
\end{figure}
\begin{figure}[tb]
\centering
\includegraphics[width=0.8\columnwidth]{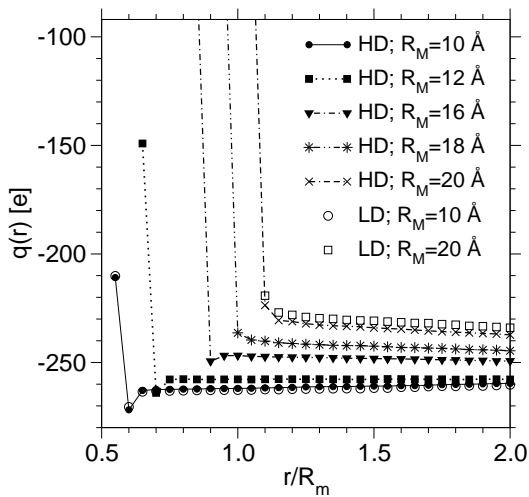}
\caption{\label{fig8}
Cumulative charge of divalent ions $q(r)$ as a function of $r/R_{\rm m}$
for different values of macroion radius $R_{\rm M}$ as indicated.  Data for the system
with $N_{\rm ct}=526$ divalent counterions and $N_{\rm co}=400$ divalent
coions is shown (the charge of the macroion is $Q=255\,e$). The distance
$r$ in the x-axis is normalised by $R_m=20$\AA~ which corresponds to the largest radius
used. }
\end{figure}
Charge inversion in the presence of multivalent counterions
has been observed in static experiments and theoretical
studies~\cite{grosberg,messina,lemay}.  Experimentally charge inversion
is detected using mainly electrophoresis. It is well established that
charge inversion is due to counterion correlations and beyond the
scope of mean--field theories. Our aim in this study is twofold: the
onset charge density $\sigma$, at which charge inversion occurs,
is determined and secondly, the extent of the Stern layer from HD simulations
is compared to that from LD simulations.

To this end, two systems of a macroion of charge $Q=255\,e$ mixed with
divalent microions are considered, one with $N_{\rm ct}=526$ divalent
counterions and $N_{\rm co}=400$ divalent coions, and the other one
with $N_{\rm ct}=176$ divalent counterions and $N_{\rm co}=50$ divalent
coions. In each case, 3 monovalent counterions are added to maintain
charge neutrality. The charge density $\sigma$ is varied by changing
the radius of the macroion $R_{\rm M}$ from $10$ to $20$\,\AA, thereby
keeping the radius $R_{\rm H}$ fixed at $10$\,\AA.  In Fig.~\ref{fig7},
the mobility $\mu$ as function of $\sigma$ is plotted for the different
systems.  Different from the case of monovalent microions, $\mu$
changes its sign in all cases, thus indicating that at high values of
$\sigma$ the macroion is moving opposite to the direction of the electric
field. Overall, at a given value of $\sigma$, the mobility of the macroion
in the presence of divalent microions is smaller than with monovalent ones.

Since the screening due to divalent microions is much more effective,
the counterion charge density is higher near the colloidal surface than 
in electrolytes with monovalent microions. This can be seen in 
Fig.~\ref{fig8}, where the system with $N_{\rm co}=400$ coions is 
considered. The cumulative charge $q(r)$ around the macroion is displayed 
for different values of $R_{\rm M}$ corresponding to different values of 
$\sigma$ at the fixed macroion charge $Q=255\,e$.  We see that for 
$R_{\rm M}\le 16$\,\AA, $q(r)$ develops a local minimum near the colloidal 
surface. This indicates charge inversion (i.e.~$|q(r)|>255\,e$) and it 
explains the appearance of negative values of $\mu$. 
%

For comparison, we have plotted LD data in Fig.~\ref{fig8} for the
largest and the smallest radii used, i.e., for $R_{\rm M}=20$\,\AA~and
$R_{\rm M}=10$\,\AA, respectively. We see that the charge profile for LD
and HD is nearly identical, which shows that the dynamic Stern layer is
determined by the static Coulomb forces. For higher salt concentration,
charge inversion is more pronounced, as seen by lower values of $\mu$ at
high $\sigma$. Both LD and HD show the phenomenon of charge inversion at
high values of $\sigma$. In particular in the HD case, charge inversion
is more pronounced for higher salt concentration. As in the studies with
monovalent counter-- and coions, for divalent microions $\mid \mu_{\rm
LD} \mid < \mid \mu_{\rm HD} \mid$ holds as well.

 
%
\section{Conclusions}
In this work, a hybrid MD/LB scheme was used to investigate the
electrophoretic mobility of a macroion in an electrolyte
solution. We have considered highly--charged macroions
($\sigma=38\,\mu$C/cm$^2$--$500\,\mu$C/cm$^2$) for which theories based
on the linearized electrokinetic equation are not applicable. For
high values of the macroion's surface charge $\sigma$, experiments on
systems with multivalent salt ions~\cite{alvarez} have shown that the
electrophoretic mobility $\mu$ can decrease with increasing $\sigma$
for $2:1$ salts whereas $\mu$ for $2:2$ salts, it shows a plateau. In
our simulations, we observe decrease of $\mu$ with $\sigma$ for both
$1:1$ and $2:2$ salt. Moreover, in contrast to  previous theoretical
studies \cite{brien}, we observe a lowering of $\mu$ with $\sigma$
even for $\kappa R_M < 3$ for monovalent salt ions.

The authors of \cite{alvarez}, attribute the lowering of $\mu$ in
the $2:1$ case to the presence of coions in the EDL. As we have
demonstrated in this work, a decreasing $\mu$ as a function of $\sigma$
requires the formation of a pronounced dynamic Stern layer of charge
$q_0$, consisting of counterions that move along the same direction as
the macroion in the presence of an electric field. The lowering of $\mu$
is directly related to the increase of the quantity $q_0/Q$ which we
call the screening charge fraction.  The absolute value of screening
charge fraction  $| q_0/Q |$ of the colloid--counterion cloud complex
increases with increasing $\sigma$, reflecting a lowering of $\mu$. Note
that the number of coions in the dynamic Stern layer is negligible.

Furthermore, we observe charge inversion for divalent counter-- and
coions, but only when we reach sufficiently high values of $\sigma$.
At very high values of $\sigma$, $\mu$ for colloidal systems with
monovalent salt ions tends to saturate, because there is no space to
add more counterions into the dynamic Stern layer. However, this is
different when multivalent salt ions and counterions are used. Then,
$\left |q_0 \right |/Q >1$ holds and $\mu$ becomes negative. The
distribution of charge in the EDL for this case as a function of distance
$r$ from the center of the colloid is non--monotic and completely
different from what is expected from standard electrokinetic theories
\cite{saville}. Though we have not carried out studies for the 2:1 salt
case, it is possible that the discrepancy in the experimental results
between the 2:1 and the 2:2 salt case arises from entropic/osmotic
forces. The larger number of coions in the 2:1 salt (compared to the 
systems with a 2:2 salt) may lead to larger osmotic forces on the 
counterions which could result in larger values of $q_0$ for the 2:1 
case. However, this issue can be resolved only after further investigations.

In the work of O'Brien and White \cite{brien}, the lowering of $\mu$ with
the $\zeta$ potential has been attributed to the distortion of the EDL.
Though the explicit measurement of the retarding force arising from the
distortion of the EDL is outside the scope of this work, we have shown
that the charge distortion within the dynamic Stern layer decreases with
increasing $\sigma$. This is indicated by the lowering of the quantities
$d/R_{\rm M}$ and $q_0 d/Q R_{\rm M}$ with increasing $\sigma$. This
observation combined with a host of other evidences presented in this
paper, indicate that accumulation of charges in the Stern layer is the
dominant mechanism for the lowering of $\mu$ as a function of $\sigma$,
providing that high values of $\sigma$ are considered.

By comparing the LB/MD (or HD) results to those from LD simulations,
we were able to elucidate the role of hydrodynamic interactions.
The structure of the Stern layer is more compact in the Langevin
simulations. Hydrodynamic interactions enhance the electrophoretic
mobility and so they have the opposite effect to electrostatic
screening. This is due to a backflow effect in the fluid which
pushes the counterions to move along the same direction as the
macroion. Hydrodynamic forces have pronounced influence in the formation
of the dynamic Stern layer, especially for low charges. An interesting
finding is that the mobility as obtained from HD relative to that from
LD exhibits an almost linear increase with $\sigma$. Further work on
the role of hydrodynamic interactions for the electrophoresis of charged
colloids is in progress.

{\bf Acknowledgment:}
We thank Burkhard D\"unweg, Vladimir Lobaskin, and Thomas Palberg for
stimulating discussions.  We acknowledge financial support by the Deutsche
Forschungsgemeinschaft (DFG) under Grants No.~HO 2231/1 and HO 2231/2,
by the SFB 625 ``Von einzelnen Molek\"ulen zu nanoskopisch strukturierten
Materialien'', and by the SFB TR6 ``Colloidal Dispersions in External
Fields''. Generous grants of computing time on the JUMP at the NIC J\"ulich 
are gratefully acknowledged.

\end{document}